\documentclass[journal=nalefd,manuscript=letter]{achemso}
\usepackage[version=3]{mhchem} 
\usepackage[T1]{fontenc} 
\usepackage{amsmath}
\usepackage{upgreek}
\usepackage{epstopdf}
\usepackage{color,soul}

\author{Evgeny M. Alexeev}
\affiliation[University of Cambridge]{Cambridge Graphene Centre, University of Cambridge, 9 JJ Thomson Avenue, CB3 0FA, Cambridge, UK}
\alsoaffiliation[University of Cambridge]{Cavendish Laboratory, University of Cambridge, JJ Thomson Avenue, Cambridge CB3 0HE, UK}
\email{ea529@cam.ac.uk}
\author{Carola M. Purser}
\affiliation[University of Cambridge]{Cavendish Laboratory, University of Cambridge, JJ Thomson Avenue, Cambridge CB3 0HE, UK}
\alsoaffiliation[University of Cambridge]{Cambridge Graphene Centre, University of Cambridge, 9 JJ Thomson Avenue, CB3 0FA, Cambridge, UK}
\author{Carmem M. Gilardoni}
\affiliation[University of Cambridge]{Cavendish Laboratory, University of Cambridge, JJ Thomson Avenue, Cambridge CB3 0HE, UK}
\author{James Kerfoot}
\author{Hao Chen}
\affiliation[University of Cambridge]{Cambridge Graphene Centre, University of Cambridge, 9 JJ Thomson Avenue, CB3 0FA, Cambridge, UK}
\author{Alisson R. Cadore}
\affiliation[University of Cambridge]{Cambridge Graphene Centre, University of Cambridge, 9 JJ Thomson Avenue, CB3 0FA, Cambridge, UK}
\alsoaffiliation[Brazilian Nanotechnology National Laboratory]{Brazilian Nanotechnology National Laboratory, Brazilian Center for Research in Energy and Materials, Sao Paulo, Brazil}
\author{B\'{a}rbara L.T. Rosa}
\affiliation[University of Cambridge]{Cambridge Graphene Centre, University of Cambridge, 9 JJ Thomson Avenue, CB3 0FA, Cambridge, UK}
\author{Matthew S. G. Feuer}
\author{Evans Javary}
\affiliation[University of Cambridge]{Cavendish Laboratory, University of Cambridge, JJ Thomson Avenue, Cambridge CB3 0HE, UK}
\alsoaffiliation[ENS PAris]{\'Ecole Normale Sup\'erieure, PSL, 5 Rue D’ulm Paris, 75005, France}
\author{Patrick Hays}
\affiliation[Arizona State University]{Materials Science and Engineering, School for Engineering of Matter,Transport and Energy, Arizona State University, Tempe, Arizona 85287, United States}
\author{Kenji Watanabe}
\affiliation[National Institute for Materials Science]{Research Center for Electronic and Optical Materials, National Institute for Materials Science, 1-1 Namiki, Tsukuba 305-0044, Japan}
\author{Takashi Taniguchi}
\affiliation[National Institute for Materials Science]{Research Center for Materials Nanoarchitectonics, National Institute for Materials Science,  1-1 Namiki, Tsukuba 305-0044, Japan}
\author{Seth Ariel Tongay}
\affiliation[Arizona State University]{Materials Science and Engineering, School for Engineering of Matter,Transport and Energy, Arizona State University, Tempe, Arizona 85287, United States}
\author{Dhiren M. Kara}
\affiliation[University of Cambridge]{Cavendish Laboratory, University of Cambridge, JJ Thomson Avenue, Cambridge CB3 0HE, UK}

\author{Mete Atat\"{u}re}
\affiliation[University of Cambridge]{Cavendish Laboratory, University of Cambridge, JJ Thomson Avenue, Cambridge CB3 0HE, UK}
\email{ma424@cam.ac.uk}
\author{Andrea C. Ferrari}
\affiliation[University of Cambridge]{Cambridge Graphene Centre, University of Cambridge, 9 JJ Thomson Avenue, CB3 0FA, Cambridge, UK}
\email{acf26@cam.ac.uk}

\title{Nature of long-lived moir\'e interlayer excitons in electrically tunable MoS$_2$/MoSe$_2$ heterobilayers}
\begin{document}

\begin{tocentry}
\includegraphics{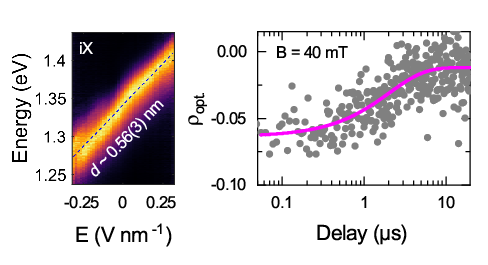}
\end{tocentry}

\begin{abstract}

Interlayer excitons in transition-metal dichalcogenide heterobilayers combine high binding energy and valley-contrasting physics with long optical lifetime and strong dipolar character. Their permanent electric dipole enables electric-field control of emission energy, lifetime, and location. Device material and geometry impacts the nature of the interlayer excitons via their real- and momentum-space configurations. Here, we show that interlayer excitons in MoS$_2$/MoSe$_2$ heterobilayers are formed by charge carriers residing at the Brillouin zone edges, with negligible interlayer hybridization. We find that the moir\'e superlattice leads to the reversal of the valley-dependent optical selection rules, yielding a positively valued g-factor and cross-polarized photoluminescence. Time-resolved photoluminescence measurements reveal that the interlayer exciton population retains the optically induced valley polarization throughout its microsecond-long lifetime. The combination of long optical lifetime and valley polarization retention makes MoS$_2$/MoSe$_2$ heterobilayers a promising platform for studying fundamental bosonic interactions and developing excitonic circuits for optical information processing.
\end{abstract}

\newpage
Van der Waals heterostructures comprising monolayer transition-metal dichalcogenides (TMDs) have emerged as a promising platform for optoelectronics\cite{Mak2016,Zhou2018,Woessner2015,Ferrari2015} and quantum technology\cite{Montblanch2023} as they combine optically addressable spin and valley degrees of freedom\cite{Wang2017a, Schaibley2016, Xiao2018} with unique tunability through the choice of material combination\cite{Novoselov2016a,Liu2019d,Catanzaro2024} and rotational alignment\cite{Wilson2021,Ciarrocchi2022,Mak2022}. TMD heterobilayers have drawn particular interest due to their ability to host interlayer excitons (iXs)\cite{Rivera2018,Regan2022} which offer lifetime approaching 200 $\mu$s\cite{Montblanch2021}, strong repulsive dipolar interaction\cite{Rivera2016,Jauregui2019}, and high sensitivity to rotational alignment\cite{Nayak2017,Kunstmann2018,Alexeev2019}, strain\cite{Kremser2020,Montblanch2021}, electric\cite{Ciarrocchi2019} and magnetic\cite{Smirnov2022} fields. Different TMD combinations give rise to iX with drastically different properties, including oscillator strength\cite{Gerber2019,Barre2022}, center-of-mass momentum\cite{Kunstmann2018,Nayak2017}, and degree of interlayer hybridization\cite{Jauregui2019,Alexeev2019}. Of the plethora of possible TMD combinations, the majority of research effort focused on MoSe$_2$/WSe$_2$\cite{Seyler2019,Tran2019,Liu2021b,Barre2022} and WS$_2$/WSe$_2$\cite{Jin2019,Stansbury2021,Tang2020b} heterobilayers. For other material combinations, key aspects of iX nature, such as real- and momentum-space configuration, remain elusive due to the complexity of the underlying physics.

In this work, we investigate iX in MoS$_2$/MoSe$_2$ heterobilayers using polarization-resolved magneto-photoluminescence spectroscopy. We find that iX photoluminescence (PL) is visible only in devices with relative twist angle less than 5$^\circ$. This indicates that the constituent iX charge carriers  reside at the edges of the Brillouin zone. We study the iX PL response to out-of-plane electric and magnetic fields and show that iX is formed by charge carriers at the $\pm$K valleys with negligible degree of interlayer hybridization. Our time- and polarization-resolved PL measurements reveal microsecond-scale retention of optically induced valley polarization, demonstrating the potential of MoS$_2$/MoSe$_2$ heterobilayers for opto-valleytronic applications. 


\begin{figure}[ht]
	\centering
	\includegraphics{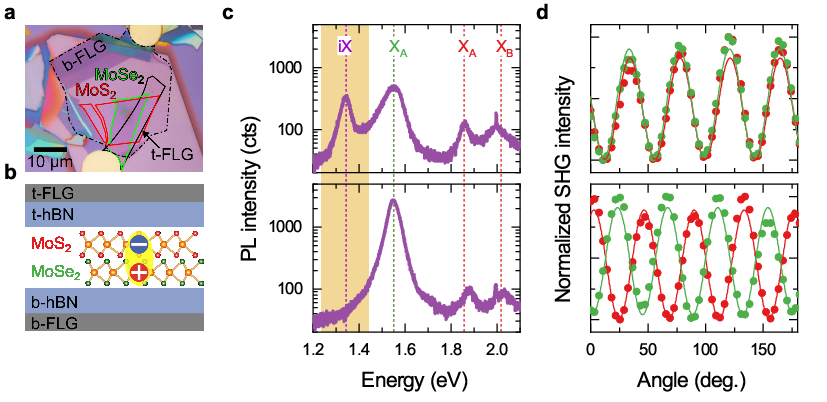}
	\caption{\textbf{iX in MoS$_2$/MoSe$_2$ heterobilayers.} (a) Optical microscope image of an electrically tunable MoS$_2$/MoSe$_2$ heterobilayer device. Monolayer MoS$_2$ and MoSe$_2$ regions are outlined in red and green, respectively. Solid (dashed) black line shows the position of top (bottom) gate composed of few-layer graphene (FLG) (b) Cross-sectional schematic of the device structure, indicating spatial configuration of iX. (c) Room-temperature PL spectra recorded in two devices with different rotational alignment. The closely rotationally aligned device (top panel, $\theta = 1^\circ$) shows intralayer MoS$_2$ B and A excitons and MoSe$_2$ A exciton peaks, as well as an iX peak appearing in a lower-energy range (highlighted in copper) that is not visible in the PL spectrum of the strongly misaligned device (bottom panel, $\theta = 28^\circ$). (d) Polarization-resolved SHG intensity recorded in the isolated monolayer (red) MoS$_2$ and (green) MoSe$_2$ regions of the two devices, confirming the twist angle of $\theta = 1^\circ$ (top) and $\theta = 28^\circ$ (bottom).}
	\label{fig:SampleImage}
\end{figure}

Figure~\ref{fig:SampleImage}a shows an optical microscope image of one of our electric-field tunable MoS$_2$/MoSe$_2$ heterobilayers, with panel b displaying its cross-section. The hexagonal boron nitride layers provide a flat and clean dielectric environment for the MoS$_2$/MoSe$_2$ heterobilayer, and the transparent few-layer graphene top and bottom gates allow optical measurements under applied out-of-plane electric field. Each of the eight devices is fabricated from micromechanically exfoliated crystals using deterministic dry mechanical transfer\cite{Castellanos-Gomez2014, Zomer2014a}, and thickness and quality of constituent layers are characterized using Raman spectroscopy\cite{Cadore2024} (see Methods and SI Fig.~S1). 
The devices offer a range of twist angles between the TMD monolayers $\theta$, enabling the investigation of iX momentum-space configuration. Figure~\ref{fig:SampleImage}c compares room-temperature PL spectra of two devices with twist angle $\theta = 1^\circ$ (top) and $\theta = 28^\circ$ (bottom). We identify $\theta$ using polarization-resolved second-harmonic generation (SHG) measurements (Fig.~\ref{fig:SampleImage}d). Both devices show PL peaks corresponding to the A exciton in MoSe$_2$ (MoSe$_2$ X$_\mathrm{A}$) at 1.55~eV and the A and B excitons in MoS$_2$ (MoS$_2$ X$_\mathrm{A}$ and X$_\mathrm{B}$) at 1.85 and 2.0~eV, respectively\cite{Cadiz2017a, Shree2020a}. Crucially, the iX PL peak at $\sim$1.3~eV is visible only in the device with $\theta = 1^\circ$. Of the eight devices  with twist angles ranging from 1$^\circ$ to 28$^\circ$ that we studied, only those with $\theta \le 5^\circ$ reveal the iX PL peak at room temperature (SI Fig.~S2), in line with a recent report\cite{Lin2023}. Thus, close rotational alignment is critical for the observation of iX in MoS$_2$/MoSe$_2$ heterobilayers.

The $\sim$3.7\% mismatch in lattice constants of MoS$_2$ and MoSe$_2$\cite{Zhu2011} eliminates twist-angle dependence of the interlayer distance as an underlying source of this behavior\cite{Constantinescu2015}. Instead, the high sensitivity of the iX PL intensity to $\theta$ indicates that the iX is formed by the charge carriers residing in valleys at the edges of the Brillouin zone. Homo- and heterobilayers where at least one of the charge carriers resides at the $\Gamma$ valley at the center of the Brillouin zone display iX PL throughout the entire $\theta$ range, as the momentum-space separation between electron and hole remains unchanged\cite{Liu2014,Kunstmann2018,Tebyetekerwa2021}. In contrast, in heterobilayers where both constituent charges reside at the edges of the Brillouin zone, momentum-space separation of electron and hole suppresses radiative recombination of iX in devices with $\theta$ away from 0 or $60^\circ$\cite{Nayak2017,Ciarrocchi2022}, consistent with our observations.			  			  
\begin{figure}[ht]
	\centering
	\includegraphics{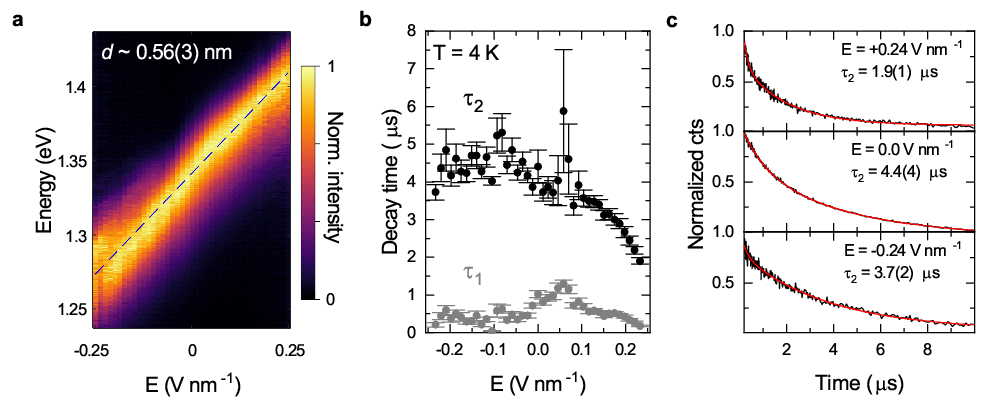}
	\caption{\textbf{Electric field tuning of iX.} (a) Normalized iX PL under applied out-of-plane electric field. iX PL energy shows a linear shift with the slope of $\sim$0.31~eV~nm~V\textsuperscript{$-1$}, corresponding to the dipole size of $ d = 0.56(3)$~nm. The dipole size closely matches the expected interlayer distance. (b) Variation of iX PL decay time as a function of applied electric field. Gray and black circles correspond to the fast ($\tau_1$) and slow ($\tau_2$) time constants, respectively. (c) PL decay acquired at +0.24, 0, and -0.24~V nm$^{-1}$ applied electric field. Red curves are bi-exponential fits to the data.}
	\label{fig:EField}
\end{figure}

We next identify the real-space configuration of iX by studying its response to applied out-of-plane electric field. Figure~\ref{fig:EField}a presents the normalized iX PL spectrum recorded as a function of electric field at 4-K temperature. The iX PL energy shifts linearly with a rate of $\sim$0.31~eV~nm~V\textsuperscript{$-1$} and can be tuned over a  144-meV range within the gate tuning limits of our device. We find an average tuning response across three devices of $\sim$0.30 eV~nm~V\textsuperscript{$-1$}, yielding an average dipole size of 0.55(3)~nm \cite{Jauregui2019,Ciarrocchi2019} (SI Fig.~S3). This value is in good agreement with the $\sim$0.6-nm separation between the layers\cite{He2014}. A similar dipole size was observed in MoSe$_2$/WSe$_2$\cite{Jauregui2019} heterobilayers with iX is formed by non-hybridized electrons and holes, while MoSe$_2$ homobilayers show a reduced dipole size of 0.26~nm due to charge-carrier hybridization\cite{Sung2020}. Comparatively, our results suggest negligible interlayer hybridization for our devices. 

Figure~\ref{fig:EField}b shows the iX PL decay time constants as a function of the applied electric field. We extract these constants from a bi-exponential fit to the time-resolved PL. Figure~\ref{fig:EField}c presents examples of PL decay trace recorded at three applied field values along with their corresponding fit curves. We observe microsecond-long iX lifetime, with a fast time constant $\tau_1 = 1.0(1)~\mu$s and a slow time constant $\tau_2 = 4.4(4)~\mu$s at zero electric field -- an order of magnitude longer than typical lifetimes of 10-100~ns reported for MoSe$_2$/WSe$_2$ heterobilayers\cite{Ciarrocchi2022}. The slow time constant in other devices ranges from 0.1~$\mu$s to 3.0~$\mu$s (SI Fig.~S4). The fast time constant is mostly field-independent, except for $\sim$0.06~V~nm\textsuperscript{$-1$}, where it increases to 1.3~$\mu$s. Shortening of $\tau_1$ for electric field away from this value is likely caused by inadvertent electrostatic doping induced by a slight asymmetry in the thicknesses of the bottom and top dielectric layers. The slow time constant $\tau_2$ shows a gradual decrease with increasing electric field, consistent with a change in radiative lifetime due to a field-induced variation of electron-hole separation. For electric field anti-parallel to the electric dipole moment of the iX, the separation between the two charge carriers is reduced, leading to an increased probability of their radiative recombination. The opposite process takes place for the parallel field alignment. That said, the PL decay time remains slow ($\tau_1 \ge 0.04~\mu$s, $\tau_2  \ge 1.9~\mu$s) throughout the entire field-tuning range. 

\begin{figure}[ht]
	\centering
	\includegraphics{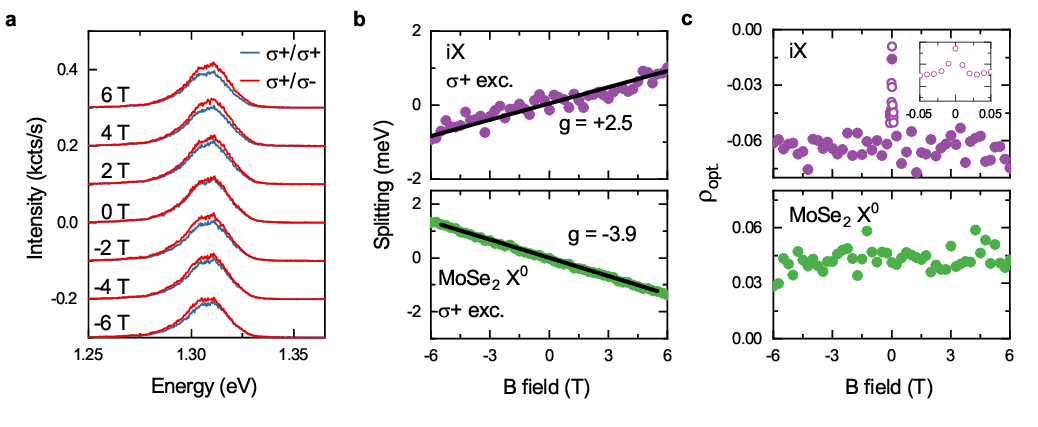}
	\caption{\textbf{Magneto-PL spectroscopy of iX.} (a) Helicity-resolved iX PL spectra recorded under applied out-of-plane magnetic field ranging from -6 to 6~T using $\sigma^+$ polarized 1.94~eV optical excitation; blue (red) curves correspond to PL with $\sigma^+$ ($\sigma^-$) polarization. (b) Energy splitting between $\sigma^+$ and $\sigma^-$ polarized PL as a function of out-of-plane magnetic field for (top) iX in heterobilayer and (bottom) neutral excitons (X$^0 $) in MoSe$_2$ monolayer regions. Land\'e g-factors extracted using linear fits are listed next to each plot. (c) Optically induced valley polarization calculated as $\rho_{\mathrm{opt}} =  (I^{++} + I^{--} - I^{+-} - I^{-+})/(I^{++} + I^{--} + I^{+-} + I^{-+})$ as a function of magnetic field for iX (top) and MoSe$_2$ X$ ^0$ (bottom), where $ I^{XY}$ is the intensity of PL with $\sigma^Y$-polarization collected under $\sigma^X$-polarized excitation. Unfilled circles in the panel and the inset are extracted from a fine scan around 0~T, showing the small-field dependence of $\rho_{\mathrm{opt}}$ for iX.}
	\label{fig:BField}
\end{figure}

We use polarization-resolved magneto-PL spectroscopy to identify valley configuration of iX. Figure~\ref{fig:BField}a shows iX PL spectra recorded using right circularly polarized ($\sigma^+$), 1.94-eV excitation as a function of applied out-of-plane magnetic field $B$ ranging from -6 to 6~T; blue (red) curves correspond to PL with $\sigma^+$ ($\sigma^-$) polarization. The iX PL remains cross-polarized with respect to the excitation laser throughout the entire magnetic field range. Two distinct mechanisms are known to give rise to this behavior in TMD heterobilayers: the directional intervalley scattering \cite{Nagler2017b} and the moir\'e-induced reversal of the valley-dependent optical selection rules \cite{Wozniak2020}. 

We identify the underlying mechanism in  MoS$_2$/MoSe$_2$ heterobilayers based on the sign of the energy splitting between $\sigma^+$ and $\sigma^-$ polarized PL under applied magnetic field. Figure~\ref{fig:BField}b is a plot of the energy splitting ($\Delta E$) as a function of $B$ for the iX in the heterobilayer region (top panel) and the neutral intralayer excitons (X$^0$) in an isolated monolayer MoSe$_2$ region (bottom panel). We define $\Delta E$ as $E_{\sigma^+} - E_{\sigma^-} = g\mu_\mathrm{B} B$, where $E_{\sigma^+}$($E_{\sigma^-}$) is the energy of the $\sigma^+$ ($\sigma^-$) polarized PL, $g$ is the effective Land\'e g-factor and $\mu_\mathrm{B}$ is the Bohr magneton. 
For X$^0$ in MoSe$_2$, we extract $ g = -3.90(2)$, consistent with previous reports\cite{Li2014a}, where the minus sign stems from valley-Zeeman interaction and valley-dependent optical selection rules for TMD monolayers\cite{Koperski2019}: $\sigma^+$-polarized light couples to optical transitions in the +K valley, which has lower energy at positive $B$. In contrast, we obtain $g = +2.50(7)$ for iX - the positive sign of $g$ shows that iX PL from the +K valley appears with $\sigma^-$ polarization, confirming the reversal of optical selection rules with respect to the monolayer case. All devices showed positive iX g-factors, with an average value of +4.5. In TMD heterobilayers, the reversal of the selection rules arises from local changes in crystal symmetry induced by the moir\'e superlattice \cite{Wozniak2020}. We observe the iX g-factors ranging from +1.0 to +8.0, likely due to the difference in the moir\'e pattern parameters arising from different twist angles and local strain in different devices (see SI Fig.~S5).

Two mechanisms can give rise to the observed PL polarization under finite magnetic field, namely the optically induced valley polarization\cite{Mak2012} and the Zeeman-shift-induced valley thermalization\cite{Cadiz2017a}. The former is limited by non-directional intervalley scattering, while the latter arises from exciton relaxation into the lower-energy valley. We calculate the degree of optically induced valley polarization independently as $\rho_{\mathrm{opt}} = \frac{I^{++} + I^{--} - I^{+-} - I^{-+}}{I^{++} + I^{--} + I^{+-} + I^{-+}}$, where $ I^{XY}$ represents the intensity of PL with $\sigma^Y$-polarization collected under $\sigma^X$--polarized excitation. Figure~\ref{fig:BField}c displays the dependence of $\rho_{\mathrm{opt}}$ on the applied magnetic field for iX and MoSe$_2$ X$ ^0$. MoSe$_2$ X$ ^0$ shows a constant PL polarization degree of $\sim$4\%, consistent with earlier reports\cite{Wang2015,MacNeill2015}. In contrast, $|\rho_{\mathrm{opt}}|$ for iX shows a distinct increase with increasing $|B|$, saturating at $\sim$6 \% above $\pm$20~mT (see inset in Fig.~\ref{fig:BField}c). This dependence is consistent across all devices, with $|\rho_{\mathrm{opt}}|$ ranging from 6\% to 14\% maximum value. In the absence of magnetic field, $|\rho_{\mathrm{opt}}|$ ranges from 0\% to 7\%. Similar sharp changes in valley polarization degree with applied magnetic field have been observed for iX in MoSe$_2$/WSe$_2$\cite{Jiang2018} and MoS$_2$/WSe$_2$\cite{Tan2021} heterobilayers, as well as intralayer excitons in WS$_2$ and WSe$_2$ monolayers\cite{Smolenski2017}. This effect was attributed to the suppression of intervalley scattering of intralayer excitons within the monolayer with dark excitonic ground state\cite{Wang2017a}. However, we observe device-specific saturation field for $\rho_{\mathrm{opt}}$ ($B_{\mathrm{sat}}$) ranging from 0.02 to 3~T (SI Fig.~S6), indicating that it is not defined by the properties of  individual monolayers, but the collective property of the assembled heterostructure.

\begin{figure}[ht]
	\centering
	\includegraphics{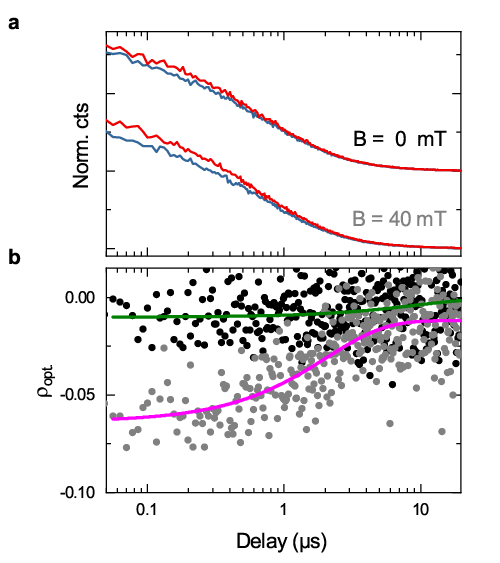}
	\caption{\textbf{Temporal evolution of iX valley polarization.} (a) Polarization-resolved iX PL decay acquired at 0 and 40~mT using $\sigma^+$ polarized excitation; blue (red) curve corresponds to PL intensity co- (cross-) polarized with the excitation laser; the data sets are offset for clarity and normalized to the intensity of co-polarized component at zero delay. (b) Time-resolved changes of $\rho_{\mathrm{opt}}$ for B = 0~mT (black) and B = 40~mT (gray). The solid curves are guides to the eye.}
	\label{fig:timeres}
\end{figure}  

Figure~\ref{fig:timeres}a presents the polarization-resolved decay of iX PL recorded at 0~mT and 40~mT in a device with $B_{\mathrm{sat}}$$\sim$200~mT. The difference in intensity for cross-co polarization allows us to monitor $|\rho_{\mathrm{opt}}|$ as a function of time. Figure~\ref{fig:timeres}b plots the time-resolved $\rho_{\mathrm{opt}}$ extracted from the iX PL decay measured at 0~mT and 40~mT; the solid curves are guides to the eye. Without magnetic field (black filled circles), iX has a low polarization degree ($|\rho_{\mathrm{opt}}| < 2\%$) throughout the measurement range. In contrast, at $B = 40$~mT (gray filled circles) $|\rho_{\mathrm{opt}}|$ starts at 6\% and shows a gradual decay towards zero, with a characteristic 1/e time of $\sim2~\mu$s. These results indicate that the loss of valley polarization for iX is governed by at least two processes occurring at different timescales: a fast intervalley relaxation with a characteristic time shorter than 10~ns (i.e., timing resolution of our measurement) dominates at zero magnetic field. This process is suppressed at 40~mT, revealing a slower, microsecond-scale relaxation.  We note that this timescale stems from both the loss of valley polarization and the decay of exciton population and therefore ultimately it provides a lower bound to the iX valley lifetime and a measure of the valley-excitonic information storage time. 
  
In conclusion, we show that the iX in MoS$_2$/MoSe$_2$ heterobilayers is formed by electrons and holes residing at the edges of the Brillouin zone, with negligible degree of interlayer hybridization. We find that iX retains its optically induced valley polarization, with the cross-polarized iX PL stemming from the moir\'e-induced reversal of selection rules. Magnetic field enhances valley-polarization retention by suppressing fast intervalley scattering. The typical magnetic field required for this ($\leq 200~$mT) is within reach of a variety of readily accessible techniques, including assembling heterostructures on magnetic substrates\cite{Gibertini2019}, or using rare-earth magnets\cite{Stern2022}. Moreover, in some devices we observe $|\rho_{\mathrm{opt}}|$ up to 7\% at zero field, allowing for magnet-free operation. The combination of microsecond-long iX PL lifetime and the retention of valley polarization offers the prospect of combining excitonic and valleytronic functionalities in a single optoelectronic device. 

\subsection{Methods}
\textbf{Sample fabrication.} 
All layers used for the fabrication of the electrically tunable MoS$_2$/MoSe$_2$ heterobilayer devices were produced by micromechanical cleavage of bulk crystals. Bulk TMD crystals were prepared by flux zone growth method\cite{Edelberg2019b} and bulk hBN crystals were grown by the temperature-gradient method\cite{Watanabe2004a}. Graphite crystals were sourced from NGS. The thickness of exfoliated crystals was estimated using optical contrast\cite{Li2013a} and confirmed using photoluminescence and Raman spectroscopy for TMDs under 532-nm (2.33-eV) laser illumination (LabRAM HR Evolution, Horiba) and atomic force microscopy for hBN (Dimension Icon, Bruker). Electrically tunable heterobilayer devices were assembled by deterministic dry mechanical transfer using polymer stamps\cite{Castellanos-Gomez2014, Zomer2014a}. The twist angle between the TMD layers was identified using polarization-resolved SHG\cite{Hsu2014} measured at room temperature using a custom-built optical setup. The SHG laser was a Chameleon Compact Optical Parametric Oscillator providing $\sim$200~fs pulses with a repetition rate of 80~MHz centered at 1320~nm. To minimize chromatic aberrations, a linearly polarized laser beam with $\sim$5-mW power was focused onto the sample using a 40x reflective objective (numerical aperture of 0.5, LMM40X-P01, Thorlabs). Polarization orientation was controlled using a super-achromatic half-wave plate (SAHWP05M-1700, Thorlabs) mounted in a motorized rotational mount. Electrical contacts to TMD layers and transparent FLG gates were created by direct laser lithography (LW-405B+, Microtech) with a positive resist (AZ5214E, MicroChemicals) followed by electron beam evaporation (PVD200Pro, Kurt J. Lesker) of 5~nm of Cr followed by 45~nm of Au. The resist excess metal layer was then lifted off by immersion in acetone and isopropanol for 30 minutes. 

\textbf{Photoluminescence measurements.} Room-temperature PL measurements were performed using LabRAM HR Evolution Raman microscope under 532-nm (2.33-eV) laser illumination. Helicity-resolved magneto-optical measurements were performed in a close-cycle bath cryostat (Attodry 1000, Attocube) equipped with a superconducting magnet at a nominal sample temperature of 4~K. Excitation and collection light pass through a home-built confocal microscope in reflection geometry, with a 0.81 numerical aperture apochromatic objective (LT-307 APO/NIR/0.81, Attocube). The PL measurements were performed using 638-nm (1.94-eV) continuous-wave excitation (MCLS1-638, Thorlabs), with incident power below 5~$\mu$W. The PL signal collected in epi-direction was isolated using a long-pass filter (FELH0700, Thorlabs) and detected by a 0.75-m spectrometer (SpectraPro 2750, Princeton Instruments) with 150~l~mm$^{-1}$ grating and a nitrogen cooled CCD camera (Spec-10, Princeton Instruments). Time-resolved measurements were performed using a single-photon avalanche photodiode (SPCM-AQRH-16-FC, Excelitas Technologies) and a time-to-digital converter (quTAU, qutools GmbH) with 81~ps timing resolution. For these measurements, the intensity of CW laser was modulated using an acousto-optic modulator (MT350-A0.12-VIS, AA Opto Electronic), producing 200-ns pulses with the 100-kHz repetition rate. A dual-channel source meter (2612B, Keithley) was used for electric field tuning.

\begin{acknowledgement}
We acknowledge funding the EU Graphene and Quantum Flagships, ERC Grants Hetero2D, GSYNCOR, GIPR, EIC Grant CHARM, EPSRC Grants EP/K01711X/1, EP/K017144/1, EP/N010345/1, and EP/L016087/1. S.A.T acknowledges primary support from DOE-SC0020653 (materials synthesis). K.W. and T.T. acknowledge support from the JSPS KAKENHI (Grant Numbers 21H05233 and 23H02052) and World Premier International Research Center Initiative (WPI), MEXT, Japan. We thank V.~Falko, C.~Faugeras, and I.~Paradisanos for fruitful discussions.
\end{acknowledgement}

\providecommand{\latin}[1]{#1}
\makeatletter
\providecommand{\doi}
  {\begingroup\let\do\@makeother\dospecials
  \catcode`\{=1 \catcode`\}=2 \doi@aux}
\providecommand{\doi@aux}[1]{\endgroup\texttt{#1}}
\makeatother
\providecommand*\mcitethebibliography{\thebibliography}
\csname @ifundefined\endcsname{endmcitethebibliography}  {\let\endmcitethebibliography\endthebibliography}{}

\end{document}


\setcounter{figure}{0}
\renewcommand{\thefigure}{S\arabic{figure}}

\begin{table}
    \centering
    \begin{tabular}{|c|c|c|c|c|c|c|c|}
        \hline
         Device \# & $\theta, ^\circ$  & $\tau_1, \mu\mathrm{s}$ & $\tau_2, \mu\mathrm{s}$ & $g$ & $\rho^0_{\mathrm{PL}}$ & $\rho^{\mathrm{sat.}}_{\mathrm{PL}}$  & $B_{\mathrm{sat.}}, T$ \\
        \hline \hline
         1 &  4&  0.03& 0.13 & +6.0  & -0.07 & -0.13 & 3\\
         \hline
         2 &  1&  1.0&4.4  & +6.7 &  -0.04&  -0.11& 1.25\\
         \hline
         3 &  3&  0.14&0.56  & +2.5 &  -0.01&  -0.07& 0.02\\
         \hline
         4 &  1&  0.33& 2.2 & +2.7 &  -0.03&  -0.10& $\le0.25$\\
         \hline
         5 &  3&  0.3&1.7  & +1.0  &  -0.01&  -0.09& $\le1$\\
         \hline
         6 &  1&  0.8& 3.0 & +8.0  &  0.0&  -0.06& 0.2\\
         \hline
    \end{tabular}
    \caption{\textbf{Devices parameters.} Summary of key parameters for the devices investigated in this study: twist angle ($\theta$), fast and slow time constant for iX PL decay ($\tau_1$ and $\tau_2$), effective iX g-factor ($g$), optically-induced valley polarization in the absence of magnetic field and at saturation ($\rho^{0}_{\mathrm{PL}}$ and $\rho^{\mathrm{sat.}}_{\mathrm{PL}}$), and saturation magnetic field for  $\rho_{\mathrm{PL}}$ ($B_{\mathrm{sat.}}$).}
    \label{tab:my_label}
\end{table}

\begin{figure}[h]
	\centering
	\includegraphics{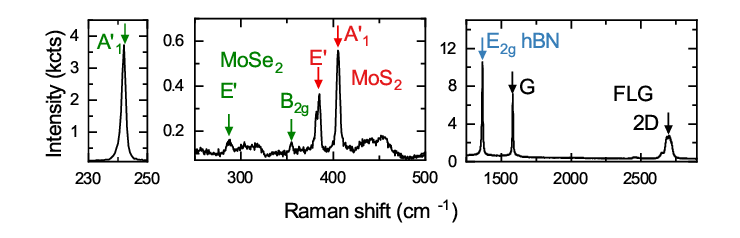}
	\caption{\textbf{Raman spectroscopy.} Raman spectra from the device, acquired at room temperature using 2.33~eV optical excitation. Dominant peaks in the 50-500~cm$^{-1}$ range correspond to in-plane $A’_1$ mode at 241.9 (405.4)~cm$^{-1}$ and out-of-plane $E’$ mode at 286.7 (384.0)~cm$^{-1}$ of MoSe$_2$ (MoS$_2$)\cite{Terrones2014}. The peak at 354.7~cm$^{-1}$ is out-of-plane $B_{2g}$ mode of MoSe$_2$, which is Raman inactive on monolayers\cite{Zeng2013}, but becomes active in few-layers and heterobilayers due to the reduction of symmetry elements\cite{Pan2022}. Higher-frequency range shows $E_{2g}$ mode at from hBN encapsulation layers at 1365~cm$^{-1}$\cite{Reich2005} and the $G$ and $2D$ Raman modes from FLG gates, at 1582 and $\sim$2697~cm$^{-1}$\cite{Ferrari2006}, respectively.}
	\label{fig:Raman}
\end{figure}

\begin{figure}[h]
	\centering
	\includegraphics{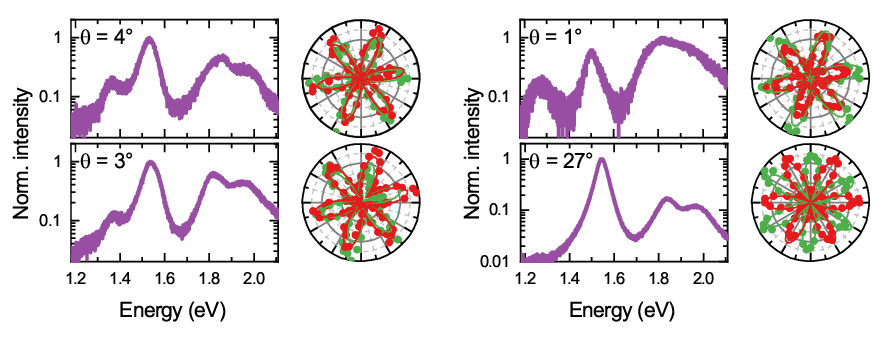}
	\caption{\textbf{Twist angle effects.} Normalized PL spectra and polar plots of polarization-resolved SHG signals acquired at room temperature in four devices with different rotational alignment between the TMD layers. Three devices with close rotational alignment ($\theta\le 5^\circ$) show iX peak in the PL spectrum, which is not present in the PL spectrum of the strongly misaligned device ($\theta \approx27^\circ$).}
	\label{fig:SHG}
\end{figure}

\begin{figure}[h]
	\centering
	\includegraphics{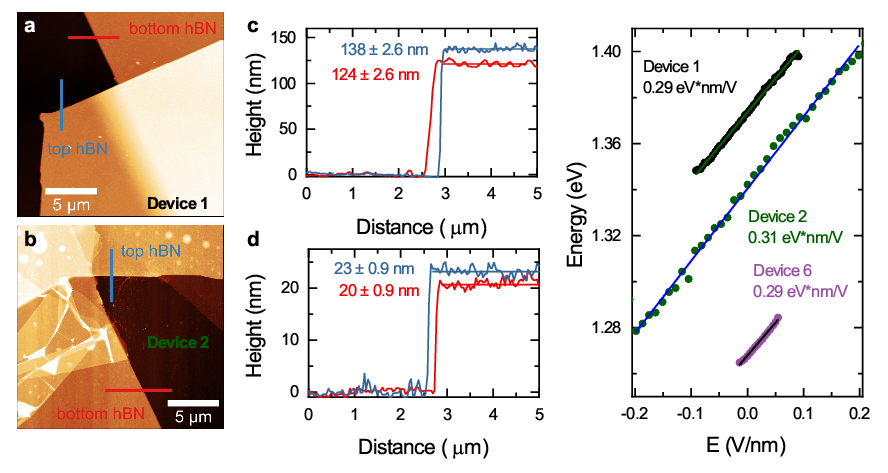}
	\caption{\textbf{Dipole size identification.} (a-c) (a and b) AFM topography images of devices 1 and 2 acquired at the edges of top and bottom hBN encapsulation layers and (c and d) cross-sections taken at the positions identified by the corresponding colour lines, showing the thickness of the layers. (e) Change of iX emission energy under applied out-of-plane electric field measured in Devices 1,2, and 6. The rate of change of $\sim$ 0.3~eV V/nm extracted from the linear fit corresponds to the dipole size of $\sim$ 0.55~nm.}
	\label{fig:AFM}
\end{figure}

\begin{figure}[h]
	\centering
	\includegraphics{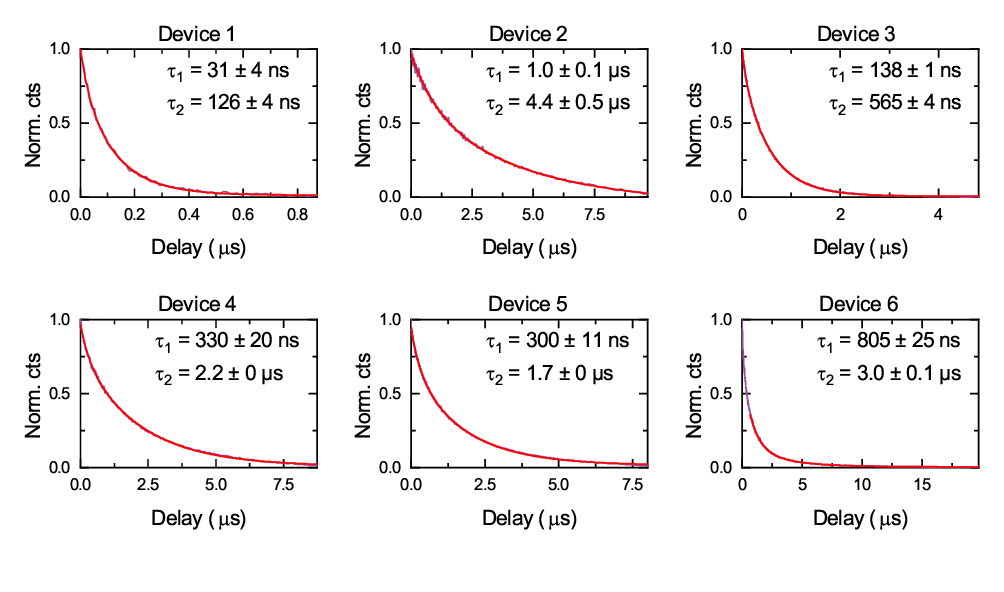}
	\caption{\textbf{iX lifetimes.} iX PL decay acquired in Devices 1-6. Solid red lines are bi-exponential fits to the data, time constants extracted from the fit are presented next to each curve.}
	\label{fig:lifetimes}
\end{figure}

\begin{figure}[h]
	\centering
	\includegraphics{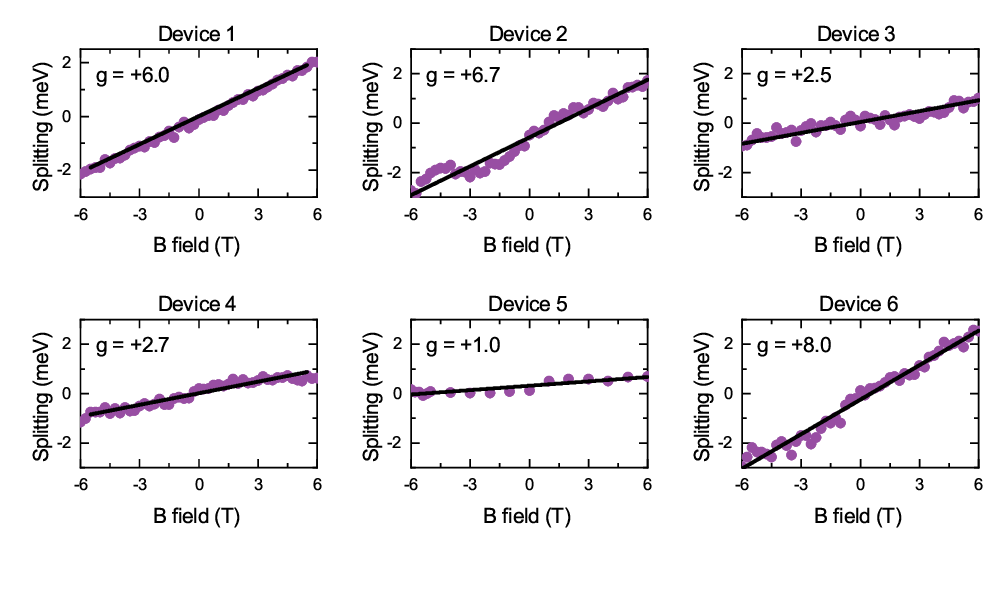}
	\caption{\textbf{iX effective g-factors.} Energy splitting between $\sigma^+$ and $\sigma^-$ polarized PL as a function of out-of-plane magnetic field for Devices 1-6. Landé g factors extracted using linear fitting are listed next to each plot.}
	\label{fig:gfactors}
\end{figure}

\begin{figure}[h]
	\centering
	\includegraphics{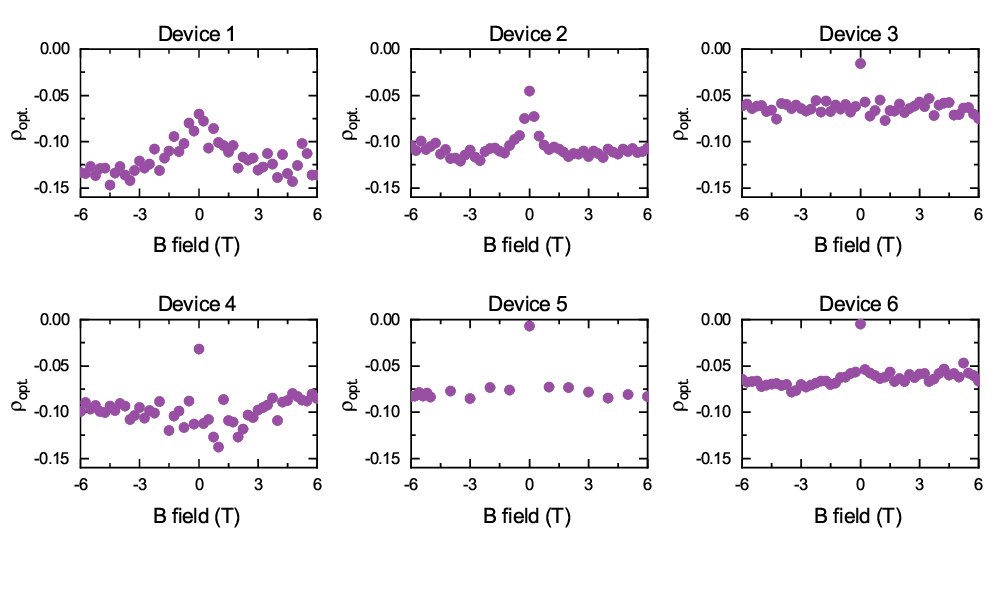}
	\caption{\textbf{Optically induced valley polarization.} Changes of optically induced iX PL valley polarization under applied magnetic field for Devices 1-6.}
	\label{fig:optpol}
\end{figure}

\clearpage

\providecommand{\latin}[1]{#1}
\makeatletter
\providecommand{\doi}
  {\begingroup\let\do\@makeother\dospecials
  \catcode`\{=1 \catcode`\}=2 \doi@aux}
\providecommand{\doi@aux}[1]{\endgroup\texttt{#1}}
\makeatother
\providecommand*\mcitethebibliography{\thebibliography}
\csname @ifundefined\endcsname{endmcitethebibliography}  {\let\endmcitethebibliography\endthebibliography}{}